\documentclass[sigconf]{acmart}

\usepackage{enumitem}
\usepackage{multirow}
\usepackage{amsmath}    
\usepackage{extarrows}  
\usepackage[table]{xcolor}

\AtBeginDocument{
  }

\newcommand{\ie}{\emph{i.e., }}
\newcommand{\eg}{\emph{e.g., }}

\newcommand{\cf}{\emph{cf. }}

\newcommand{\model}{\textsl{MealRec} }
\newcommand{\modelwospace}{\textsl{MealRec}}

\setcopyright{acmlicensed}
\copyrightyear{2018}
\acmYear{2018}
\acmDOI{XXXXXXX.XXXXXXX}

\acmConference[ACM KDD '26]{32nd SIGKDD Conference on Knowledge Discovery and Data Mining}{August 09-13, 2026}{Jeju, Korea}

\acmISBN{978-1-4503-XXXX-X/2018/06}

\begin{document}
\begin{sloppypar}

\title{\textsl{MealRec}: Multi-granularity Sequential Modeling via Hierarchical Diffusion Models for Micro-Video Recommendation}

\author{Xinxin Dong}
\email{dongxinxin@nudt.edu.cn}
\affiliation{
  \institution{National University of Defense Technology}
  \city{Changsha}
  \country{China}
}

\author{Haokai Ma}
\email{haokai.ma1997@gmail.com}
\authornotemark[1]
\thanks{*Corresponding author}
\affiliation{
 \institution{National University of Singapore}
 \city{Singapore}
 \country{Singapore}}

\author{Yuze Zheng}
\email{zhengyuzezzz@gmail.com}
\affiliation{
  \institution{National University of Defense Technology}
  \city{Changsha}
  \country{China}
}

\author{Yongfu Zha}
\email{zhayongfu@nudt.edu.cn}
\affiliation{%
  \institution{National University of Defense Technology}
  \city{Changsha}
  \country{China}
}

\author{Yonghui Yang}
\email{yyh.hfut@gmail.com}
\affiliation{%
  \institution{National University of Singapore}
  \city{Singapore}
  \country{Singapore}}

\author{Xiaodong Wang}
\email{xdwang@nudt.edu.cn}
\affiliation{%
  \institution{National University of Defense Technology}
  \city{Changsha}
  \country{China}}

\renewcommand{\shortauthors}{Trovato et al.}

\begin{abstract}

Micro-video recommendation aims to capture user preferences from the collaborative and context information of the interacted micro-videos, thereby predicting the appropriate videos. This target is often hindered by the inherent noise within multimodal content and unreliable implicit feedback, which weakens the correspondence between behaviors and underlying interests. While conventional works have predominantly approached such scenario through behavior-augmented modeling and content-centric multimodal analysis, these paradigms can inadvertently give rise to two non-trivial challenges: preference-irrelative video representation extraction and inherent modality conflicts. To address these issues, we propose a \underline{M}ulti-granularity s\underline{e}quential modeling method vi\underline{a} hierarchica\underline{l} diffusion models for micro-video \underline{Rec}ommendation (\textbf{\underline{\modelwospace}}), which simultaneously considers temporal correlations during preference modeling from intra- and inter-video perspectives. Specifically, we first propose Temporal-guided Content Diffusion (TCD) to refine video representations under intra-video temporal guidance and personalized collaborative signals to emphasize salient content while suppressing redundancy. To achieve the semantically coherent preference modeling, we further design the Noise-unconditional Preference Denoising (NPD) to recovers informative user preferences from corrupted states under the blind denoising. Extensive experiments and analyses on four micro-video datasets from two platforms demonstrate the effectiveness, universality, and robustness of our \modelwospace, further uncovering the effective mechanism of our proposed TCD and NPD. The source code and corresponding dataset will be available upon acceptance.
\end{abstract}



\maketitle
\section{Introduction}

\begin{figure}[t]
    \centering
    \includegraphics[width=0.98\linewidth]{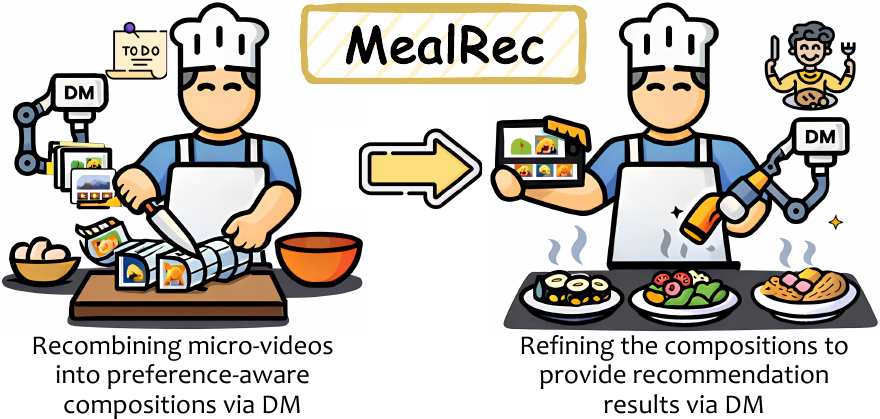}
    \vspace{-0.3cm}
    \caption{Illustration of the proposed \model as a chef preparing a meal: it first leverages the ``kitchenware'' (DM) to sequentially slice and recombine ``ingredients'' (micro-videos) into ``preferences-aware compositions'' (video representations), and then applies another ``tool'' (DM) to professionally refine these ``compositions'', ultimately serving the delicious ``meal'' (recommendation results).}
    \vspace{-0.4cm}
    \label{fig:motivation_graph}
\end{figure}

With the proliferation of short-form video platforms, micro-videos have become a dominant mode of online content consumption. Effective micro-video recommendation is crucial for boosting user engagement and increasing platform stickiness. Unlike conventional recommendation tasks~\cite{CIERec,RealHNS,NS4RS}, micro-video recommendation must simultaneously capture the dynamic collaborative dependencies across the interacted video sequence of each user and the visual evolution within each video's frame sequence, posing additional challenges for both academia and industry~\cite{ni2025content,shang2023learning,xu2025mutual,zhao2025multi}.

Considering the specific challenges, existing video recommendation endeavors largely fall into two paradigms: behavior-augmented modeling and content-centric multimodal analysis. The former predominantly leverages the auxiliary behavioral signals of users (\ie watch time, comments, and shares) to minutely model their preferences from the collaborative perspective. However, this paradigm heavily depends on the availability of platform-specific logs, making it difficult to transfer to more general micro-video recommendation settings that only can access implicit behavioral sequences and video contents. For the latter, prior work typically models user preferences from the modality perspective via modality-specific preference capturing~\cite{wei2019mmgcn}, segment-level interest modeling~\cite{he2025short}, or multimodal foundation models adapting with parameter-efficient fine-tuning strategies~\cite{fu2025efficient}. However, these approaches typically summarize each video in the corpus into a static content representation, overlooking the temporal dynamics across frames and failing to effectively align content with user preferences, thereby yielding suboptimal performance in recommendation.

Inspired by the spatio-temporal dynamics modeling and uncertainty denosing of Diffusion Models (DMs), many researcher have investigated their effectiveness in both video understanding~\cite{ho2022video}, multimodal recommendation~\cite{lu2025dmmd4sr} and sequential recommendation~\cite{mao2025distinguished} tasks. Riding this momentum, recent studies have moved toward the more challenging multimodal sequential recommendation task. Here, DMMD4SR~\cite{lu2025dmmd4sr} employs multi-level DMs with uncertainty-guided fusion to mitigate interest-agnostic noise in pre-trained multimodal features, while M$^3$BSR~\cite{cui2025multi} integrates iterative denoising with cross-modal fusion for preference modeling. Notably, diffusion models have demonstrated effectiveness in video understanding, as iterative denoising over spatiotemporal representations captures inter-frame dependencies and dynamic visual patterns. These success promote us a potential solution: \emph{Can we jointly unify the strengths of DMs in multimodal sequential recommendation and video understanding to improve micro-video recommendation?} 

Despite being reasonable, it poses non-trivial challenges: preference-irrelative video representation extraction and inherent modality conflicts. Therefore, we propose \modelwospace, a hierarchical diffusion framework to formulate micro-video recommendation as a multi-granularity sequential modeling process. As illustrated in Figure~\ref{fig:motivation_graph}, drawing an analogy to a chef's cooking process, the proposed \model introduces Temporal-guided Content Diffusion (TCD) and Noise-unconditional Preference Denoising (NPD) to reframe micro-video recommendation. Specifically, we first propose TCD module at the intra-video level to simultaneously incorporate the frame-lavel temporal guidance and personalized collaborative signals into micro-video content extraction, thereby mitigating the preference-irrelative video representation extraction issue. Inspired by recent findings that a DMs' noise level can be inferred from corrupted representations without explicit noise scheduling~\cite{sun2025noise}, we propose a NPD module at the inter-video level. Here, NPD performs multimodal fusion via blind denoising, which directly recovers informative preference representations from corrupted states under the guidance of video and textual representations without timestep–conditioned $\epsilon$-prediction, thereby seamlessly mitigating inherent modality conflicts.

We have conducted extensive experiments on four micro-video datasets from two diverse platforms to demonstrate the superiority of our \modelwospace. We also provide various ablation study, visualization, robustness analysis on noise, backbones, important parameters and complexity to proves the effectiveness of the proposed TCD and NPD. The contributions can be summarized as follows:
\begin{itemize}[leftmargin=*, topsep=0.2pt,parsep=0pt]
\item We propose \modelwospace, a hierarchical diffusion framework for micro-video recommendation that jointly model the intra-video temporal dynamics and inter-video preference evolution. To the best of our knowledge, we are first to explore multi-granularity sequential modeling issue in micro-video recommendation.
\item The proposed \model couples TCD to refine preference-guided video representations with tailored U-Net estimator from the intra-video frames and NPD to reconstruct user interests from inter-video interactions via blind denoising.
\item Extensive experiments on four real-world datasets demonstrates the effectiveness, universality, robustness, and lightweight characteristics of our \model.
\end{itemize}

\section{Related Work}
\subsection{Micro-Video Recommendation}

Micro-videos inherently encompass rich multimodal information, including visual frames, audio tracks, and textual descriptions. Effectively leveraging these heterogeneous modalities has become a central challenge in micro-video recommendation. MMGCN~\cite{wei2019mmgcn} pioneered the application of graph convolutional networks to multimodal recommendation by constructing modality-specific user-item interaction graphs. M3CSR~\cite{chen2024multi} preprocesses multimodal features via pretrained encoders and obtains trainable category IDs, while MLLM-MSR~\cite{ye2025harnessing} leverages MLLMs for multimodal item summarization and user preference inference. However, these methods typically treat videos as atomic units at video level, overlooking the fine-grained interest drift that occurs at different moments within individual videos at the segment level. To address this limitation, another line of work focuses on modeling fine-grained user interests within individual videos. FRAME~\cite{shang2023learning} decomposes videos into clips and employs graph convolutional networks to model user-clip interactions, capturing interest variations across different 
temporal segments from both positive and negative perspectives. SINE~\cite{pan2023understanding} identifies dominant sub-interests by jointly optimizing positive and passive-negative feedback at the segment level through multi-task learning, enabling the modeling of shifting user attention within videos. SegRec~\cite{he2025short} advances segment-level preference modeling by fusing heterogeneous modalities to capture dynamic interest patterns. Nevertheless, while these methods capture temporal interest dynamics, they do not fully exploit rich content information within videos. In contrast, our approach bridges this gap by modeling user preferences at the frame level, integrating content-aware multimodal understanding with fine-grained interest modeling, thereby enabling more precise preference representation.

\subsection{Sequential Recommendation}

Sequential recommendation models users' dynamic preferences by capturing temporal dependencies in interaction histories. Early methods relied on Markov Chains~\cite{rendle2010factorizing} and RNNs~\cite{hidasi2015session}, while Transformer-based models have become the dominant paradigm with self-attention. SASRec~\cite{kang2018self} introduced unidirectional self-attention for sequence modeling, and BERT4Rec~\cite{sun2019bert4rec} adopted bidirectional attention with masked item prediction. Later studies further improved sequential modeling with factors such as time intervals~\cite{li2020time}, personalization~\cite{wu2020sse}, and contrastive learning~\cite{xie2022contrastive}.

Recent work incorporates multimodal information to enrich item representations beyond sparse interactions. MMSR~\cite{hu2023adaptive} proposes adaptive multimodal fusion, and MP4SR~\cite{zhang2024multimodal} introduces multimodal pre-training with contrastive objectives. Beyond multimodal fusion, LLM-based recommendation has also attracted attention for personalization, latent reasoning, and conversational strategy optimization~\cite{zhao2025nextquill,zhang2025reinforced,zhao2025reinforced}. However, existing multimodal sequential recommendation methods typically treat multimodal features as auxiliary side information, simply concatenating or fusing them with ID embeddings without fully exploiting modality semantics. Moreover, they primarily rely on coarse item-level tags rather than fine-grained content cues, which limits their ability to fully characterize users' dynamic preferences on multimodal content; related exploration in video recommendation remains limited.

\subsection{Diffusion Models for Recommendation}

Diffusion models have recently emerged as a powerful generative paradigm in recommender systems due to their superior capability in modeling complex data distributions and denoising corrupted signals~\cite{TriCDR,SeeDRec,MCDRec,HorizonRec}. DiffRec~\cite{wang2023diffusion} first applies the denoising diffusion probabilistic model to learn the generative process of user-item interactions, demonstrating advantages over traditional generative models in handling noisy interactions. Subsequent studies have extended diffusion models to various recommendation scenarios.
For collaborative filtering, prior work has explored diffusion modeling by leveraging high-order connectivity and graph-structured signals to better capture user–item interaction patterns~\cite{hou2024collaborative,zhu2024graph}. In the context of sequential recommendation, DiffuRec~\cite{li2023diffurec} represents items as distributions rather than fixed vectors to capture users' diverse preferences, while DiffuASR~\cite{liu2023diffusion} employs diffusion models for data augmentation to alleviate data sparsity and long-tail user problems. PDRec~\cite{ma2024plug} proposes a plug-in diffusion framework that infers users' dynamic preferences via a time-interval diffusion model. More recently, diffusion models have also been introduced into multimodal sequential recommendation to jointly model interaction sequences and multimodal item content~\cite{lu2025dmmd4sr}. Despite these advances, existing diffusion-based recommenders often model ID-based interactions or coarse item-level features, falling short of leveraging fine-grained multimodal semantics. In particular, effectively incorporating intra-video semantic and temporal cues into the diffusion process for sequential recommendation remains largely unexplored.

\section{Preliminary}
\subsection{Diffusion Models}
\label{sec:pre_diffusion}
Diffusion models are generative models that learn to synthesize data by reversing a progressive noise corruption process. In this section, we provide a brief overview of the forward diffusion process, the reverse denoising process, and the training objective.

\noindent
\textbf{Forward Process.} The forward process progressively corrupts data by adding Gaussian noise over $T$ timesteps. Starting from an initial data sample $\mathbf{x}_0 \sim q(\mathbf{x}_0)$, the forward process is defined as a Markov chain:
\begin{equation}
    q(\mathbf{x}_t | \mathbf{x}_{t-1}) = \mathcal{N}(\mathbf{x}_t; \sqrt{1-\beta_t}\mathbf{x}_{t-1}, \beta_t \mathbf{I}),
\end{equation}
where $\{\beta_t\}_{t=1}^{T}$ is a predefined variance schedule. A key property of this process is that we can sample $\mathbf{x}_t$ at any arbitrary timestep $t$ directly from $\mathbf{x}_0$ in closed form:
\begin{equation}
    q(\mathbf{x}_t | \mathbf{x}_0) = \mathcal{N}(\mathbf{x}_t; \sqrt{\bar{\alpha}_t}\mathbf{x}_0, (1-\bar{\alpha}_t)\mathbf{I}),
\end{equation}
where $\bar{\alpha}_t = \prod_{s=1}^{t}(1-\beta_s)$. This allows efficient sampling via $\mathbf{x}_t = \sqrt{\bar{\alpha}_t}\mathbf{x}_0 + \sqrt{1-\bar{\alpha}_t}\boldsymbol{\epsilon}$ with $\boldsymbol{\epsilon} \sim \mathcal{N}(\mathbf{0}, \mathbf{I})$.

\noindent
\textbf{Reverse Process.} The reverse process aims to recover the original data from noise by learning a denoising estimator $f_\theta(\cdot)$. Given the noised input $\mathbf{x}_t$ and optional conditioning information $\mathbf{c}$, the estimator can be designed to predict either the original data $\mathbf{x}_0$ or the noise component $\boldsymbol{\epsilon}$:
\begin{equation}
    \hat{\mathbf{x}}_0 = f_\theta(\mathbf{x}_t, \mathbf{c}, t) = f_\theta(\sqrt{\bar{\alpha}_t}\mathbf{x}_0 + \sqrt{1-\bar{\alpha}_t}\boldsymbol{\epsilon}, \mathbf{c}, t),
\end{equation}
where $\mathbf{c}$ denotes task-specific contextual guidance. During inference, samples are generated by iteratively denoising from $\mathbf{x}_T \sim \mathcal{N}(\mathbf{0}, \mathbf{I})$ to obtain $\mathbf{x}_0$.

\noindent
\textbf{Training Objective.} Depending on the prediction target, diffusion models can be trained with different objectives. For direct data prediction, we minimize the Mean Squared Error (MSE) between the original and estimated data:
\begin{equation}
    \mathcal{L}_{\mathbf{x}_0} = \mathbb{E}_{\mathbf{x}_0, \boldsymbol{\epsilon}, t}\left[\|\mathbf{x}_0 - f_\theta(\sqrt{\bar{\alpha}_t}\mathbf{x}_0 + \sqrt{1-\bar{\alpha}_t}\boldsymbol{\epsilon}, \mathbf{c}, t)\|^2\right].
\end{equation}

\subsection{Base Recommender}
\label{sec:base_recommender}

Since user preferences are inherently time-varying and viewing behaviors are observed in chronological order, we adopt a sequential recommender as the backbone. Specifically, we use SASRec~\cite{kang2018self} as the base encoder, which builds on the Transformer architecture and leverages self-attention to model complex item-to-item transitions in user behavior sequences. Given an input embedding matrix $\mathbf{D} \in \mathbb{R}^{L \times d}$ formed by combining item embeddings with positional encodings, where $L$ is the sequence length and $d$ is the latent dimension, SASRec applies multi-head self-attention to compute contextualized representations:

\begin{equation}
    \text{Attention}(\mathbf{Q}, \mathbf{K}, \mathbf{V}) = \text{Softmax}\left(\frac{\mathbf{Q}\mathbf{K}^T}{\sqrt{d}}\right)\mathbf{V},
\end{equation}

where $\mathbf{Q} = \mathbf{D}\mathbf{W}^Q$, $\mathbf{K} = \mathbf{D}\mathbf{W}^K$, and $\mathbf{V} = \mathbf{D}\mathbf{W}^V$ denote the query, key, and value matrices obtained through learnable projection matrices $\mathbf{W}^Q, \mathbf{W}^K, \mathbf{W}^V \in \mathbb{R}^{d \times d}$. A position-wise feed-forward network is subsequently applied to enhance representation capacity. The module yields $L$ contextualized $d$-dimensional states, each encoding the representation at its corresponding position.

To demonstrate the generality of our framework, we evaluate three representative sequential recommendation backbones: SASRec~\cite{kang2018self}, CL4SRec~\cite{xie2022contrastive}, and TedRec~\cite{xu2024tedrec}, spanning self-attention, contrastive learning, and multimodal fusion. These results verify that our approach is agnostic to the base encoder. Detailed analyses are provided in Section~\ref{sec:robustness_analysis}.

\begin{figure*}[t]
    \includegraphics[width=0.95\linewidth]{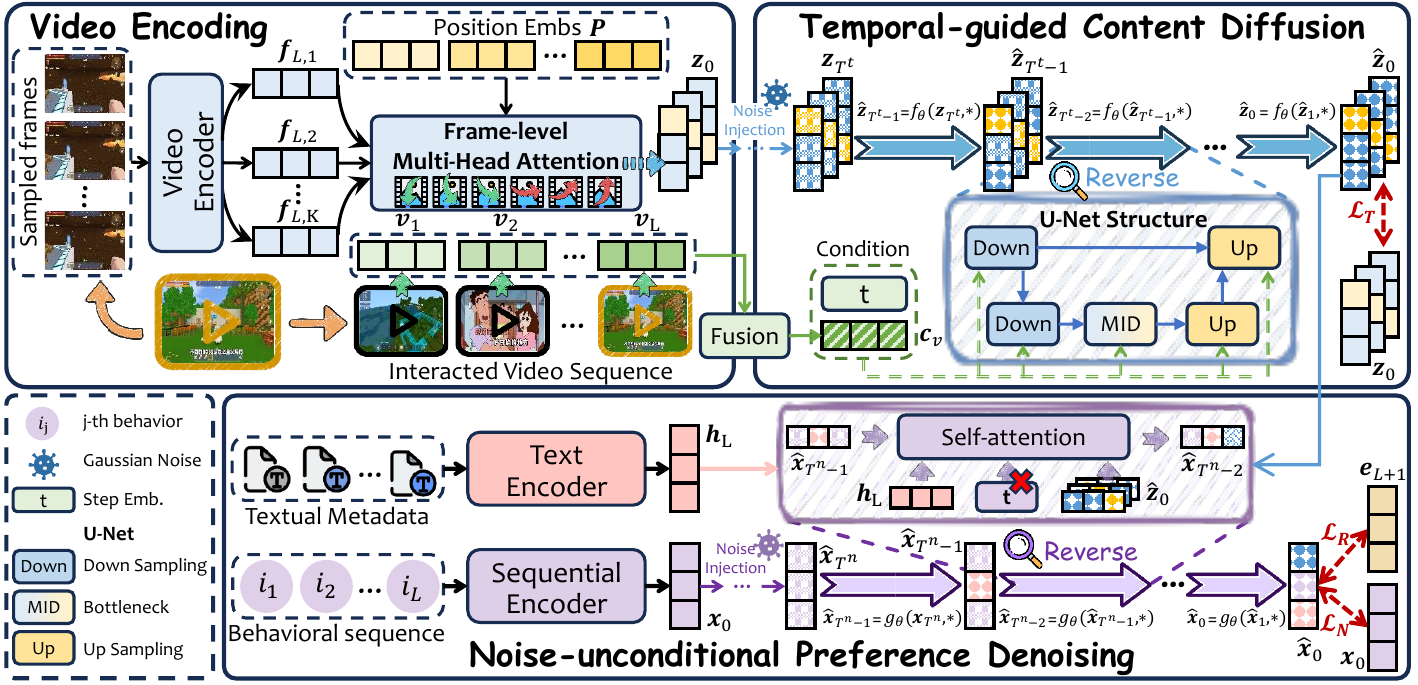}
    \vspace{-0.3cm} 
    \caption{Overall structure of our proposed \modelwospace.}
    \vspace{-0.4cm}
    \label{fig:overall_stracture}
\end{figure*}

\section{\includegraphics[scale=0.02]{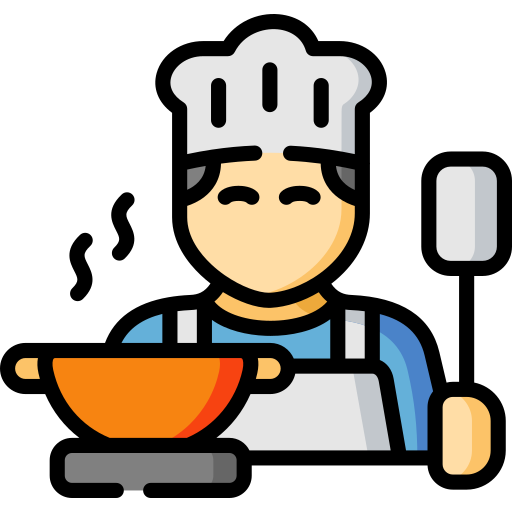}~MealRec}

\subsection{Problem Formulation}

In this work, we focus on the micro-video sequential recommendation task. Let $\mathcal{U}$ and $\mathcal{I}$ denote the sets of users and items, respectively. For each user $u \in \mathcal{U}$, we represent their historical interaction sequence as $\mathcal{S}_u = [i_1, i_2, \dots, i_L]$, where $i_k \in \mathcal{I}$ represents the $k$-th interacted item and $L$ is the sequence length. Each item $i$ is characterized by a multimodal feature tuple $\mathcal{M}_i = (\mathbf{v}_i, \mathbf{F}_i, \mathbf{h}_i)$, where $\mathbf{v}_i \in \mathbb{R}^{d_v}$ is a global visual embedding summarizing the overall video content, $\mathbf{F}_i = [\mathbf{f}_{i,1}, \mathbf{f}_{i,2}, \dots, \mathbf{f}_{i,K}] \in \mathbb{R}^{K \times d_v}$ is a sequence of $K$ sampled frame-level features capturing fine-grained temporal dynamics, and $\mathbf{h}_i \in \mathbb{R}^{d_h}$ is the textual embedding extracted from titles and descriptions. Given the user's interaction sequence $\mathcal{S}_u$ and the associated multimodal features, our objective is to predict the next item $i_{L+1}$ that best aligns with the user's evolving preferences.

\vspace{-0.25cm}
\subsection{Overall Framework}

To address noise at multiple temporal scales in micro-video recommendation, we propose \model, a hierarchical denoising framework for sequential modeling. As illustrated in Figure~\ref{fig:overall_stracture}, \model~adopts a hierarchical denoising architecture with two components operating at different granularities. At the \textit{intra-video} level, TCD refines frame-level representations by modeling temporal dependencies within videos, producing preference-informed video features. These refined features are then incorporated into user modeling. At the \textit{inter-video} level, NPD performs denoising over interaction sequences to recover clean user preference representations from noisy behavioral histories. The final recommendations are generated by ranking candidate items according to the denoised user preference representations.
    
\vspace{-0.1cm}
\subsection{Temporal-guided Content Diffusion}
\label{sec:tcd}
Existing methods typically compress videos into static embeddings, overlooking fine-grained temporal dependencies across frames. To address this limitation, TCD explicitly models frame-level temporal dynamics to capture the evolving visual content within micro-videos. The TCD module integrates temporal dependency modeling with diffusion-based denoising to refine frame-sequence representations, conditioned on the user's historical viewing behavior. As illustrated in Figure~\ref{fig:overall_stracture}, TCD consists of two components: (1) \textit{intra-video temporal dependency modeling}, which captures temporal relationships across frames, and (2) \textit{diffusion-based feature denoising and refinement}, which progressively improves representation quality under preference guidance.

\subsubsection{Intra-Video Temporal Dynamics Modeling.}
To capture intra-video temporal structure, TCD models frame-level dependencies within the micro-video of the user's most recent interaction $i_L$ (the $L$-th item in $\mathcal{S}_u=[i_1,\dots,i_L]$), which provides the most up-to-date evidence of current user intent. We represent this micro-video as a sequence of sampled frame features $\mathbf{F}_{L}=[\mathbf{f}_{L,1},\dots,\mathbf{f}_{L,K}] \in \mathbb{R}^{K\times d_v}$. To preserve temporal order, we augment the frame features with learnable positional embeddings $\mathbf{P}\in\mathbb{R}^{K\times d_v}$ and apply a temporal attention encoder:
\begin{equation}
\mathbf{z}_0 = \mathrm{TempAttnEnc}(\mathbf{F}_{L} + \mathbf{P}),
\end{equation}

where $\mathbf{z}_0 \in \mathbb{R}^{K \times d_v}$ denotes the temporally contextualized frame representations. To incorporate the user's evolving preferences from their interaction history, we condition the subsequent diffusion refinement on a recency-aware visual context vector $\mathbf{c}_v$ aggregated via exponentially decayed weights:
\begin{equation}
\alpha_k = \frac{\exp(-\gamma (L-k))}{\sum_{j=1}^{L}\exp(-\gamma (L-j))}, \qquad
\mathbf{c}_v = \sum_{k=1}^{L}\alpha_k\, \mathbf{v}_k,
\end{equation}
where $\gamma>0$ controls the decay rate and $\{\mathbf{v}_k\}_{k=1}^{L}$ are the global visual features of historical items. The context vector $\mathbf{c}_v$ serves as the conditioning signal, guiding the diffusion process to emphasize recent viewing patterns.

\subsubsection{Diffusion-based Temporal Context Denoising.}
We adopt the standard forward diffusion process in Section~\ref{sec:pre_diffusion} to corrupt $\mathbf{z}_0$ and obtain noisy latents $\mathbf{z}_t$ under the predefined schedule $\{\beta_t\}_{t=1}^{T}$. Increasing $t$ progressively destroys the temporal structure in $\mathbf{z}_0$.
To reverse the corruption under preference guidance, we parameterize a conditional denoising network $f_\theta$ using a U-Net architecture adapted for temporal sequences. At each reverse step $t \in \{T, T\!-\!1, \dots, 1\}$, the network predicts the clean latent:
\begin{equation}
\hat{\mathbf{z}}_0^{(t)} = f_\theta(\mathbf{z}_t, t, \mathbf{c}_v),
\label{eq:clean_estimator}
\end{equation}
where $\mathbf{c}_v$ is injected via cross-attention to provide user preference guidance. Using this prediction, we compute the denoised latent at the previous step via the reverse posterior:
\begin{equation}
\mathbf{z}_{t-1} = \boldsymbol{\mu}_\theta(\mathbf{z}_t, t, \mathbf{c}_v) + \sigma_t \boldsymbol{\epsilon}, \quad \boldsymbol{\epsilon} \sim \mathcal{N}(\mathbf{0}, \mathbf{I}),
\label{eq:reverse_step}
\end{equation}
where $\boldsymbol{\mu}_\theta$ is derived from $\hat{\mathbf{z}}_0^{(t)}$ following the standard DDPM formulation. This iterative refinement proceeds as:
\begin{equation}
\mathbf{z}_T \!\xlongrightarrow{f_\theta(\mathbf{z}_T, T, \mathbf{c}_v)}\! \mathbf{z}_{T-1} \!\xlongrightarrow{f_\theta(\mathbf{z}_{T-1}, T\!-\!1, \mathbf{c}_v)}\! \cdots \!\xlongrightarrow{f_\theta(\mathbf{z}_1, 1, \mathbf{c}_v)}\! \hat{\mathbf{z}}_0,
\label{eq:reverse_chain}
\end{equation}
yielding preference-aligned frame representations $\hat{\mathbf{z}}_0 \in \mathbb{R}^{K \times d_v}$ for downstream user modeling.
Accordingly, we optimize $f_\theta$ with a diffusion reconstruction objective that promotes faithful recovery of the clean latent under $\mathbf{c}_v$. Formally, we minimize:

\begin{equation}
\label{eq:loss_tcd}
\mathcal{L}_{\mathrm{T}} = 
\mathbb{E}_{\mathbf{z}_0, t}\!\left[\left\| \mathbf{z}_0 - f_\theta(\mathbf{z}_t, t, \mathbf{c}_v) \right\|^2\right].
\end{equation}

\subsection{Noise-unconditional Preference Denoising }
\label{sec:npd}

Although intra-video temporal modeling captures dynamics within individual videos, it overlooks modeling dependencies across the user’s interaction sequence. In practice, behavioral logs are often contaminated by noisy actions (e.g., accidental clicks and exploratory views), which can obscure the underlying preference signal.  
Motivated by recent findings that corrupted high-dimensional observations can reveal noise characteristics~\cite{sun2025noise}, we propose a timestep-free estimator that performs blind preference recovery using multimodal content priors.

We model the user's interaction sequence $\mathcal{S}_u = [i_1, \dots, i_L]$ as a noisy manifestation of their underlying preferences. Let $\mathbf{E} \in \mathbb{R}^{|\mathcal{I}| \times d}$ denote the item embedding matrix. We apply a sequential encoder to the embedded sequence $[\mathbf{e}_1, \dots, \mathbf{e}_L]$ to obtain the initial preference representation:
\begin{equation}
    \mathbf{x}_0 = \mathrm{SeqEnc}([\mathbf{e}_1, \dots, \mathbf{e}_L]) \in \mathbb{R}^d.
\end{equation}
Following Section~\ref{sec:pre_diffusion}, we apply the standard forward diffusion process to corrupt $\mathbf{x}_0$ and obtain the noisy latent $\mathbf{x}_t$ at step $t$, where the noise level is governed by a predefined schedule.

To provide rich semantic guidance for preference reconstruction, we condition the denoising process on two complementary content signals from the latest interacted item $i_L$: (1) the TCD-refined visual representation $\hat{\mathbf{z}_0} \in \mathbb{R}^{K\times d_v}$, which encapsulates preference-aligned temporal dynamics, and (2) the textual feature $\mathbf{h}_L \in \mathbb{R}^{d_h}$ extracted from titles and descriptions.

We parameterize the denoising network $g_\theta$ using a self-attention mechanism that enables each modality to attend to others, facilitating semantic alignment and noise identification. Crucially, $g_\theta$ operates \emph{without} access to the diffusion timestep $t$, compelling the model to infer noise characteristics purely from the semantic consistency across modalities. 

\vspace{-0.3cm}
\begin{equation}
    \mathbf{x}_T \!\xlongrightarrow{g_\theta(\mathbf{x}_T, \mathbf{c}_m)}\! \mathbf{x}_{T-1} \!\xlongrightarrow{g_\theta(\mathbf{x}_{T-1}, \mathbf{c}_m)}\! \mathbf{x}_{T-2} \!\xlongrightarrow{g_\theta(\mathbf{x}_{T-2}, \mathbf{c}_m)}\! \cdots \!\xlongrightarrow{g_\theta(\mathbf{x}_1, \mathbf{c}_m)}\! \hat{\mathbf{x}}_0,
\label{eq:npd_reverse_chain}
\end{equation}

Notably, $g_\theta$ is timestep-agnostic and therefore performs blind denoising by relying solely on cross-modal semantic consistency between $\hat{\mathbf{z}}_0$ and $\mathbf{h}_L$. Moreover, the multimodal condition $\mathbf{c}_m$ remains fixed throughout the reverse process, anchoring the denoising trajectory to the content of $i_L$ and mitigating representation drift.

Accordingly, we optimize $g_\theta$ with a reconstruction objective that encourages accurate recovery of $\mathbf{x}_0$ from noisy inputs under the fixed multimodal condition:
\begin{equation}
\label{eq:loss_npd}
\mathcal{L}_{\mathrm{N}} = 
\mathbb{E}_{\mathbf{x}_0, t}\!\left[\left\| \mathbf{x}_0 - g_\theta(\mathbf{x}_t, \hat{\mathbf{z}}_0, \mathbf{h}_L) \right\|^2\right].
\end{equation}

\subsection{Optimization Objective}

We optimize a discriminative next-item prediction objective via cross-entropy over the item set $\mathcal{I}$:
\begin{equation}
\label{eq:loss_rec}
\mathcal{L}_{\mathrm{R}} = -\sum_{(u,i)\in\mathcal{D}} 
\log \frac{e^{\hat{y}_{ui}}}{\sum_{k \in \mathcal{I}} e^{\hat{y}_{uk}}}.
\end{equation}

\label{sec:optimization}
We train the framework end-to-end with a hybrid objective that combines a discriminative ranking loss and two denoising reconstruction losses:
\begin{equation}
\label{eq:total_loss}
\mathcal{L}_{\mathrm{total}} =
\mathcal{L}_{\mathrm{R}}
+ \lambda_{\mathrm{T}} \mathcal{L}_{\mathrm{T}}
+ \lambda_{\mathrm{N}} \mathcal{L}_{\mathrm{N}},
\end{equation}
where $\lambda_{\mathrm{T}}$ and $\lambda_{\mathrm{N}}$ balance the denoising terms. 
$\mathcal{L}_{\mathrm{R}}$, $\mathcal{L}_{\mathrm{T}}$, and $\mathcal{L}_{\mathrm{N}}$ are defined in Eq.~\eqref{eq:loss_rec}, Eq.~\eqref{eq:loss_tcd}, and Eq.~\eqref{eq:loss_npd}, respectively.

\begin{table}[t]
\centering
\caption{The detailed statistics of the datasets.}
\vspace{-0.2cm}
\label{tab:dataset_stats}
\resizebox{\columnwidth}{!}{%
\begin{tabular}{lccccc}
\toprule
Dataset & \#Users & \#Items & \#Interactions & Avg. Length & Sparsity \\
\midrule
Microlens-small  & 8,694  & 8,265   & 50,838  & 5.85  & 99.93\% \\
Microlens-big    & 86,073 & 18,700  & 597,104 & 6.94  & 99.96\% \\
Shortvideo-small & 3,460  & 84,192  & 240,843 & 69.61 & 99.92\% \\
Shortvideo-big   & 7,584  & 122,845 & 533,381 & 70.33 & 99.94\% \\
\bottomrule
\end{tabular}}
\vspace{-0.3cm}
\end{table}

\begin{table*}[t]
\centering
\caption{Performance comparison on our \model against ten baselines across four micro-video datasets from two platforms. $^{*}$ denotes significant improvements of \model over the baselines (\emph{p} $\textless$ 0.01 with paired t-tests).}
\vspace{-0.2cm}
\label{tab:performance_comparison_extended}
\resizebox{\textwidth}{!}{%
\begin{tabular}{lcccccccccccccccc}
\toprule
\multirow{2}{*}{\textbf{Algorithm}} & \multicolumn{4}{c}{\textbf{Microlens-small}} & \multicolumn{4}{c}{\textbf{Microlens-big}} & \multicolumn{4}{c}{\textbf{Shortvideo-small}}  & \multicolumn{4}{c}{\textbf{Shortvideo-big}} \\
\cmidrule(lr){2-5} \cmidrule(lr){6-9} \cmidrule(lr){10-13} \cmidrule(lr){14-17}
 & \textbf{H@10} & \textbf{N@10} & \textbf{H@20} & \textbf{N@20} & \textbf{H@10} & \textbf{N@10} & \textbf{H@20} & \textbf{N@20} 
 & \textbf{H@10} & \textbf{N@10} & \textbf{H@20} & \textbf{N@20} & \textbf{H@10} & \textbf{N@10} & \textbf{H@20} & \textbf{N@20} \\
\midrule

SASRec (18)      & 0.0781 & 0.0459 & 0.1000 & 0.0515 & 0.0714 & 0.0363 & 0.1082 & 0.0465 & 0.0121 & 0.0073 & 0.0171 & 0.0085 & 0.0133 & 0.0064 & 0.0224 & 0.0087 \\  
CL4SRec (22)    & 0.0789 & 0.0478 & 0.1020 & 0.0536 & 0.0735 & 0.0382 & 0.1122 & 0.0479  & 0.0142 & 0.0082 & 0.0214 & 0.0100  & 0.0178 & 0.0085 & 0.0280 & 0.0110 \\
SSDRec (24)     & 0.0710 & 0.0377 & 0.0942 & 0.0436 & 0.0558 & 0.0265 & 0.0867 & 0.0343 & 0.0090 & 0.0056 & 0.0116 & 0.0062 & 0.0127 & 0.0077 & 0.0161 & 0.0085  \\
DiQDiff (25)    & 0.0614 & 0.0399 & 0.0682 & 0.0416  & 0.0734 & 0.0403 & 0.1091  & 0.0492  & 0.0083 & 0.0049 & 0.0147 & 0.0065  & 0.0129 & 0.0065 & 0.0217 & 0.0087  \\
\midrule

MoRec (23)      & 0.0681 & 0.0322 & 0.1028 & 0.0410 & 0.0540 & 0.0265 & 0.0893 & 0.0354  & 0.0140 & 0.0080 & 0.0216 & 0.0099  & 0.0145 & 0.0072 & 0.0198 & 0.0086  \\
TedRec (24)     & \underline{0.0837} & 0.0463 & 0.1056 & 0.0518 & 0.0743 & 0.0387 & 0.1113 & 0.0481  & 0.0153 & 0.0078 & \underline{0.0249} & 0.0102 & \underline{0.0185} & 0.0089 & \underline{0.0301} & \underline{0.0118}  \\     
IISAN (24)      & 0.0793 & 0.0397 & 0.1024 & 0.0457 & 0.0759 & 0.0390 & 0.1147 & 0.0488  & 0.0154 & 0.0086 & 0.0228 & 0.0104  & 0.0140 & 0.0079 & 0.0200 & 0.0094  \\
DMMD4SR (25)    & 0.0805 & 0.0431  & 0.1011  & 0.0483 & \underline{0.0796} & 0.0385 & \underline{0.1182} & 0.0482 & 0.0139 & 0.0068 & 0.0234 & 0.0091 & 0.0182 & \underline{0.0089} & 0.0281 & 0.0113   \\ \midrule     
MMGCN (19)      & 0.0369 & 0.0150  &  0.0774 & 0.0253 & 0.0202 & 0.0093 & 0.0348 & 0.0130 & 0.0116 & 0.0053 & 0.0173 & 0.0068 & 0.0117 & 0.0066 & 0.0193 & 0.0084  \\
IISAN-Verse (25)    &0.0830 & \underline{0.0478} & \underline{0.1139} & \underline{0.0536} & 0.0769 & \underline{0.0414} & 0.1114 & \underline{0.0501} & \underline{0.0175} & \underline{0.0089} & 0.0236 & \underline{0.0104} & 0.0154 & 0.0083 & 0.0222 & 0.0100 \\\midrule
\rowcolor{blue!6} \textbf{MealRec (Ours)} & \textbf{0.0897*} & \textbf{0.0498*} & \textbf{0.1195*} & \textbf{0.0573*} & \textbf{0.0809} & \textbf{0.0431*} & \textbf{0.1225*} & \textbf{0.0536*} & \textbf{0.0194*} & \textbf{0.0090} & \textbf{0.0292*} & \textbf{0.0115*} & \textbf{0.0207*} & \textbf{0.0104*} & \textbf{0.0335*} & \textbf{0.0136*} \\
\rowcolor{blue!6} \textit{Rel. Imp.} & \textit{7.17\%} & \textit{4.18\%} & \textit{4.92\%} & \textit{6.90\%} & \textit{1.63\%} & \textit{4.11\%} & \textit{3.64\%} & \textit{6.99\%} & \textit{10.86\%} & \textit{1.12\%} & \textit{17.27\%} & \textit{10.58\%} & \textit{11.89\%} & \textit{16.85\%} & \textit{11.30\%} & \textit{15.25\%} \\ \bottomrule
\end{tabular}%
}
\vspace{-0.3cm}
\end{table*}

\subsection{Discussion}
\subsubsection{Complexity Analysis}
We further analyze the computational complexity of training \textsc{MealRec} and discuss its comparative efficiency with respect to standard diffusion-based recommender models. The training cost of MealRec is dominated by sequence modeling and fine-grained visual reconstruction. Let $L$, $d$, and $K$ denote the sequence length, hidden dimension, and the number of sampled video frames, respectively. The sequential encoder and the attention-based NPD estimator both have complexity $\mathcal{O}(L^{2}d)$, since self-attention is used to capture temporal dependencies in the user history. The U-Net-based TCD module performs temporal convolution over $K$ frames per item, leading to $\mathcal{O}(LKd^{2})$ complexity. Since TCD frame processing is fully parallelizable across items and NPD is trained with single-timestep sampling, the overall training complexity is $\mathcal{O}(L^{2}d + LKd^{2})$, making MealRec computationally comparable to standard multimodal sequential recommenders.

\subsubsection{Comparison with Existing Approaches}
Existing micro-video recommendation methods mainly fall into two paradigms. (i) \emph{Pre-trained feature-based} approaches extract video features offline and treat them as fixed inputs for sequential modeling. While effective, they cannot adapt visual representations to user-specific interests and often ignore fine-grained intra-video temporal cues. (ii) \emph{End-to-end feature learning} approaches optimize visual encoders jointly with recommendation objectives, which typically entails considerable training-time computation and storage overhead. Compared with these paradigms, our framework is designed to alleviate both issues. Specifically, TCD performs preference-guided diffusion refinement to produce temporally contextualized and preference-aligned visual representations, while NPD recovers user preferences through iterative denoising guided by multimodal content priors. This design decouples preference reconstruction from noisy interactions and yields robust user representations for recommendation.

\section{Experiments} 
We conduct extensive experiments and analyses to answer the following research questions:\\
\noindent
\textbf{RQ1}: How does \model  perform against the state-of-the-art baselines in micro-video recommendation?  (\cf Section~\ref{sec:performance_comparison}) \\
\noindent
\textbf{RQ2}:  How does each component proposed in \model impact its recommendation performance? (\cf Section~\ref{sec:ablation_study}) \\
\noindent
\textbf{RQ3}: How does \model contribute to the video representations and user preference modeling? (\cf Section~\ref{sec:Visualization}) \\
\noindent
\textbf{RQ4}: Is \model still effective with other base sequential encoders (\cf Section~\ref{sec:UniversalityAnalysis}) \\
\noindent
\textbf{RQ5}: How does \model perform under varying noise levels, hyper-parameter settings and how does its computational complexity compare with competitive baselines? (\cf Section~\ref{sec:robustness_analysis}) \\

\subsection{Experimental Setups}

\subsubsection{Datasets}

We conduct experiments on four real-world micro-video datasets from two platforms: Microlens-small and Microlens-big~\cite{ni2023content}, as well as Shortvideo-small and Shortvideo-big~\cite{shang2025large}. To improve data reliability, we apply 4-core filtering and construct chronologically ordered interaction sequences for each user. We follow the leave-one-out setting, where the last and second last interactions are used for testing and validation, respectively. Key dataset statistics are summarized in Table~\ref{tab:dataset_stats}.

\subsubsection{Baselines}

To assess the effectiveness of \modelwospace, we compare it against diverse baselines, which are organized into three distinct categories: sequential recommendation methods (\ie SASRec~\cite{kang2018self}, CL4SRec~\cite{xie2022contrastive}, SSDRec~\cite{zhang2024ssdrec}, and DiQDiff~\cite{mao2025distinguished}), multimodal sequential recommendation methods (\ie MoRec~\cite{yuan2023go}, TedRec~\cite{xu2024tedrec}, IISAN~\cite{fu2024iisan}, and DMMD4SR~\cite{lu2025dmmd4sr}), and video recommendation methods (\ie MMGCN~\cite{wei2019mmgcn} and IISAN-Verse~\cite{fu2025efficient}).

\subsubsection{Implementation Details}
All experiments are conducted on a single NVIDIA A100 GPU with Python 3.10.18. For visual inputs, we follow the frame extraction strategy in~\cite{ni2023content} by uniformly sampling frames at 1 fps from the central portion of each video, yielding five frames per video. These frames are encoded using the pre-trained VideoMAE~\cite{tong2022videomae} to obtain both frame-level features and a global visual representation for multimodal modeling. For text, we extract textual features using BERT~\cite{devlin2019bert} on the MicroLens datasets and GloVe~\cite{pennington2014glove} on the Shortvideo datasets.

We instantiate \model using SASRec, CL4SRec, and TedRec as alternative sequential backbones while keeping the rest of the architecture unchanged. For a fair comparison, we adopt the same backbone configuration for all variants. Specifically, the sequential encoder comprises two self-attention blocks with two attention heads, and the maximum sequence length is set to 10. We train the model with Adam using a learning rate of $1\times10^{-3}$ and a batch size of 2048. In the TCD module, we set the time-decay factor to $\gamma=0.9$. During training, we sample the diffusion step $t$ uniformly from $\{5,10,15,20,25\}$, and use the same number of diffusion steps for both TCD and NPD to reduce hyper-parameter tuning, i.e., $T^{t}=T^{n}=T$. The loss weights $\lambda_{\mathrm{T}}$ and $\lambda_{\mathrm{N}}$ are tuned via grid search over $\{0.01, 0.05, 0.1, 0.3, 0.5, 0.7, 0.9\}$.

\begin{figure*}[t]
    \includegraphics[width=0.98\linewidth]{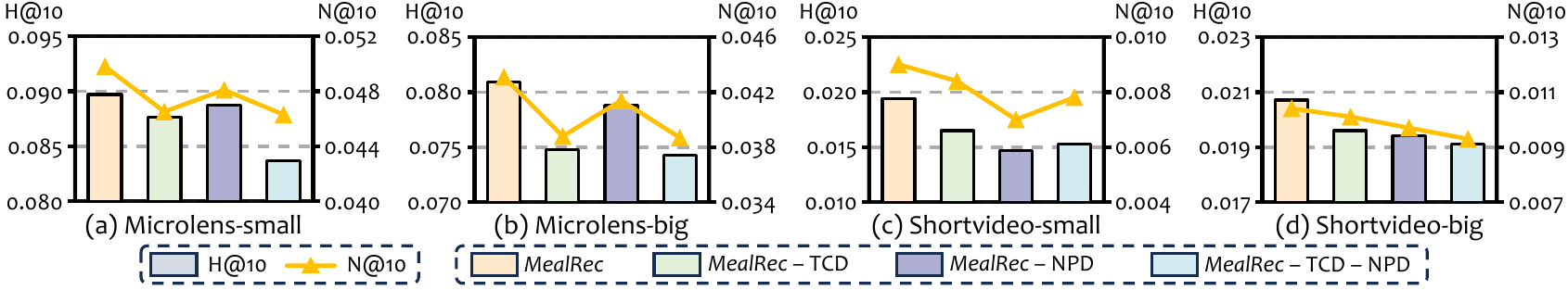}
    \vspace{-0.3cm}
    \caption{Ablation analysis of \model across four datasets. Generally, each component within our \model is effective.}
    \vspace{-0.3cm}
    \label{fig:ablation_study}
\end{figure*}

\begin{figure*}[!t]
    \includegraphics[width=0.98\linewidth]{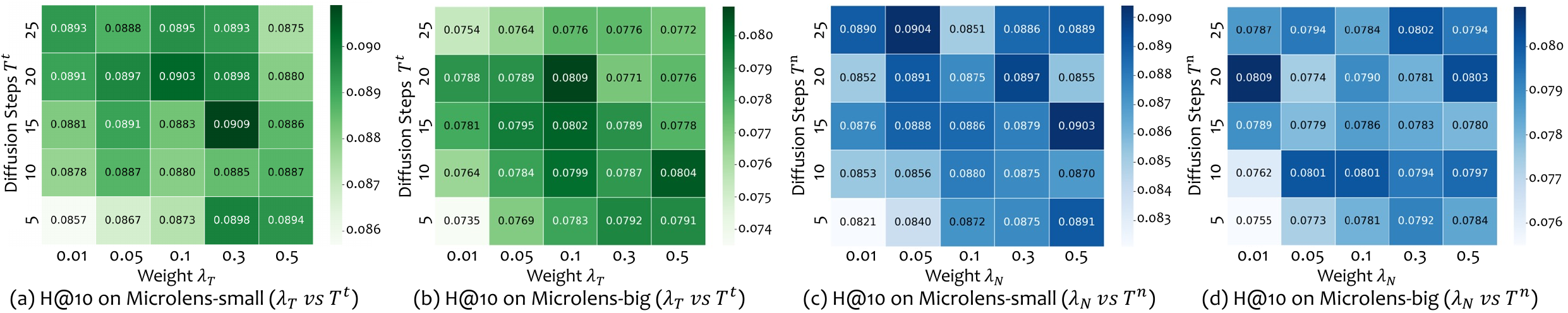}
    \vspace{-0.3cm}
    \caption{Parameter sensitivity analysis about (a) \& (b): The comparison between the loss weight $\lambda_{\mathrm{T}}$ and diffusion step $T^{\mathrm{t}}$ of TCD; (c) \& (d): The comparison between the loss weight $\lambda_{\mathrm{T}}$ and diffusion step $T^{\mathrm{t}}$ of NPD. Here, darker colors denote higher scores.}
    \vspace{-0.4cm}
    \label{fig:Parameter_Sensitivity_TCD}
\end{figure*}

\subsection{Performance Comparison (RQ1)}
\label{sec:performance_comparison} 
Table~\ref{tab:performance_comparison_extended} reports the performance comparison on four datasets in terms of Hit Ratio (H@k) and NDCG (N@k), where k$\in\!\{10,20\}$. Here, the best and second-best results are highlighted in \textbf{bold} and \underline{underlined}, respectively. Overall, \model consistently achieves the best results across all datasets and metrics, demonstrating its effectiveness for micro-video recommendation. We attribute these gains to its unified modeling of fine-grained intra-video dynamics and sequence-level preference denoising.

Compared with \textbf{sequence-only} baselines (\eg SASRec, CL4SRec, SSDRec, and DiQDiff), \model achieves consistent improvements by incorporating video and textual content. These gains highlight the value of multimodal semantics for micro-video recommendation, where fine-grained visual details and descriptive cues provide informative signals of user interests. By incorporating content-aware representations, \model better differentiates items with similar visual themes and improves the accuracy of identifying truly preference-matched videos.

Compared with \textbf{multimodal sequential} baselines (\eg MoRec, TedRec, IISAN, and DMMD4SR), \model achieves further improvements by exploiting fine-grained intra-video content information. Existing methods primarily rely on item-level representations and thus often overlook the rich temporal dynamics within individual videos. Our results underscore the importance of modeling intra-video structure beyond coarse-grained features for micro-video recommendation. By incorporating frame-level temporal representations, \model better disentangles user interests and improves recommendation accuracy.

Compared with \textbf{video-oriented} baselines (\eg MMGCN and IISAN-Verse), the results highlight the importance of jointly modeling sequential dependencies and intra-video temporal structure. \model consistently outperforms these methods, indicating improved robustness and generalization. These gains mainly stem from hierarchical diffusion, which denoises preference representations while preserving micro-video temporal dynamics.

\subsection{Ablation Study (RQ2)}
\label{sec:ablation_study}  

To investigate the effectiveness of each proposed module, we conduct ablation studies on TedRec to quantify the impact of the diffusion components. As shown in Figure~\ref{fig:ablation_study}, we evaluate several variants to examine the role of each component. We can observe that: 1) \textbf{MealRec-TCD} replaces the refined video representation with the initial feature extracted by the video encoder (i.e., disabling TCD-based visual diffusion), which tests whether diffusion-based visual refinement is necessary. The results show that TCD yields consistent gains on most datasets, highlighting its effectiveness in capturing user viewing interests from fine-grained video dynamics. The slight degradation on Shortvideo-small suggests that user sequences may be noisier in this setting, indicating the need for further denoising of interaction sequences. 2) \textbf{MealRec-NPD} removes the NPD diffusion module and instead applies an MLP to fuse multimodal features. The resulting performance degradation across all datasets indicates that NPD plays an important role in alleviating noise in user interaction sequences and stabilizing preference evolution. \noindent 3) \textbf{MealRec-TCD-NPD} disables both diffusion modules, leading to the largest performance drop among all variants. This result demonstrates that visual diffusion and preference denoising are complementary, and jointly modeling fine-grained video dynamics and denoised preference sequences is crucial to performance improvement in \modelwospace.

\subsection{Robustness Analysis (RQ5)}
\label{sec:robustness_analysis} 

\begin{figure}[!t]
    \includegraphics[width=1\linewidth]{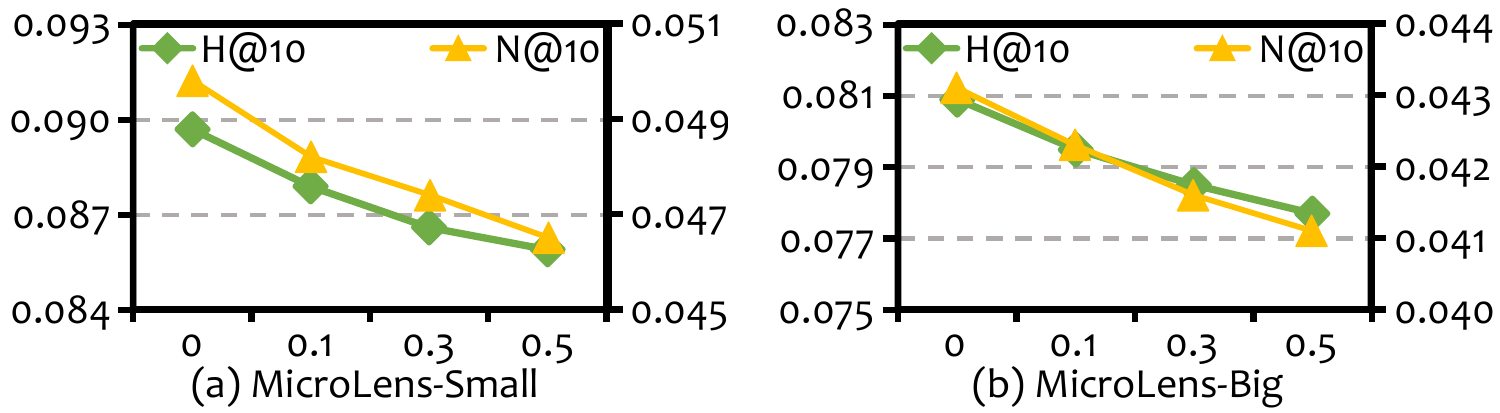}
    \vspace{-0.5cm}
    \caption{Performance of \model under different levels of Gaussian noise ($\sigma$) added to the pre-extracted visual features.}
    \vspace{-0.4cm}
    \label{fig:Noisy_robustness}
\end{figure}

\begin{figure*}[!t]
    \includegraphics[width=0.98\linewidth]{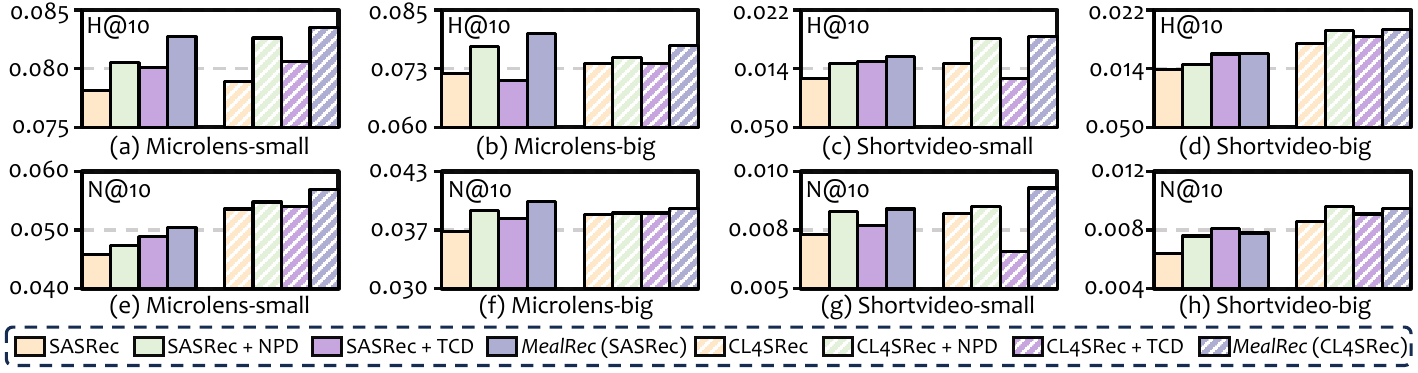}
    \vspace{-0.3cm}
    \caption{Universality analysis of \modelwospace. We show the results of all the ablation variants of \model on SASRec and CL4SRec.}
    \vspace{-0.4cm}
    \label{fig:Universityr_Analysis}
\end{figure*}

\subsubsection{Robustness Analysis on Diverse Noisy Level} 

To analyze the robustness of MealRec under perturbations in video frame representations, we inject standard Gaussian noise with different magnitudes (0, 0.1, 0.3, and 0.5) into the video embeddings produced by the video encoder on the Microlens-small and Microlens-big datasets. Figure~\ref{fig:Noisy_robustness} illustrates the trends of H@10 and N@10 under varying noise levels. As the noise intensity gradually increases, both H@10 and N@10 exhibit only slight declines without noticeable degradation. This robustness is partly due to the hierarchical diffusion in MealRec, which leverages recency-aware user guidance to refine visual representations and complement collaborative information.

\subsubsection{Parameter Sensitivity Analysis}

Figure~\ref{fig:Parameter_Sensitivity_TCD} reveals consistent trends across datasets and diffusion components. For both TCD and NPD, \model exhibits stable performance over a wide range of hyperparameter settings, indicating low sensitivity to the choice of loss weights and diffusion steps. In particular, moderate loss weights combined with intermediate to larger diffusion steps consistently achieve strong performance, suggesting a favorable balance between reconstruction strength and denoising depth. Moreover, the relatively smooth performance landscapes across datasets demonstrate that the proposed framework is robust to hyperparameter variations, which is desirable for practical deployment where extensive tuning is often impractical.

\begin{table}[!t]
    \centering
    \caption{Computational complexity comparison between TedRec (backbone), IISAN-Versa (SOTA baseline) and our \modelwospace. Here ``\#Tra.'', ``\#Eval.'' and ``\#Mem.'' denote training time and evaluation time per epoch, and the memory usage.}
    \vspace{-0.1cm}
    \label{tab:complexity_analysis}

    \resizebox{\linewidth}{!}{
        \begin{tabular}{c|c|c|ccc}
            \toprule 
            \textbf{Dataset} & \textbf{Algorithms} & \textbf{H@20} & \textbf{\#Tra. (s)} & \textbf{\#Eval. (s)} & \textbf{\#Mem. (MB)} \\
            \midrule 
            \multirow{3}{*}{\begin{tabular}[c]{@{}c@{}}MicroLens \\ -small\end{tabular}} 
             & TedRec      &0.1056 & 1.23 & 1.47 & 2,716 \\
             & IISAN-Versa & 0.1139 & 180.53 & 12.63 & 40,894 \\
             & \modelwospace & 0.1195& 1.87 & 2.31 & 3,280 \\
            \midrule 
            \multirow{3}{*}{\begin{tabular}[c]{@{}c@{}}MicroLens \\ -big\end{tabular}}
             & TedRec      & 0.0893 & 16.85 & 14.71 & 3,574 \\
             & IISAN-Versa & 0.1114 & 2493.07 & 107.40 & 41,336 \\
             & \modelwospace & 0.1225 & 24.94 & 22.08 & 4,012 \\
             
            \bottomrule 
        \end{tabular}
    }
\vspace{-0.4cm}
    
\end{table}

\begin{figure}[!t]

    \includegraphics[width=0.98\linewidth]{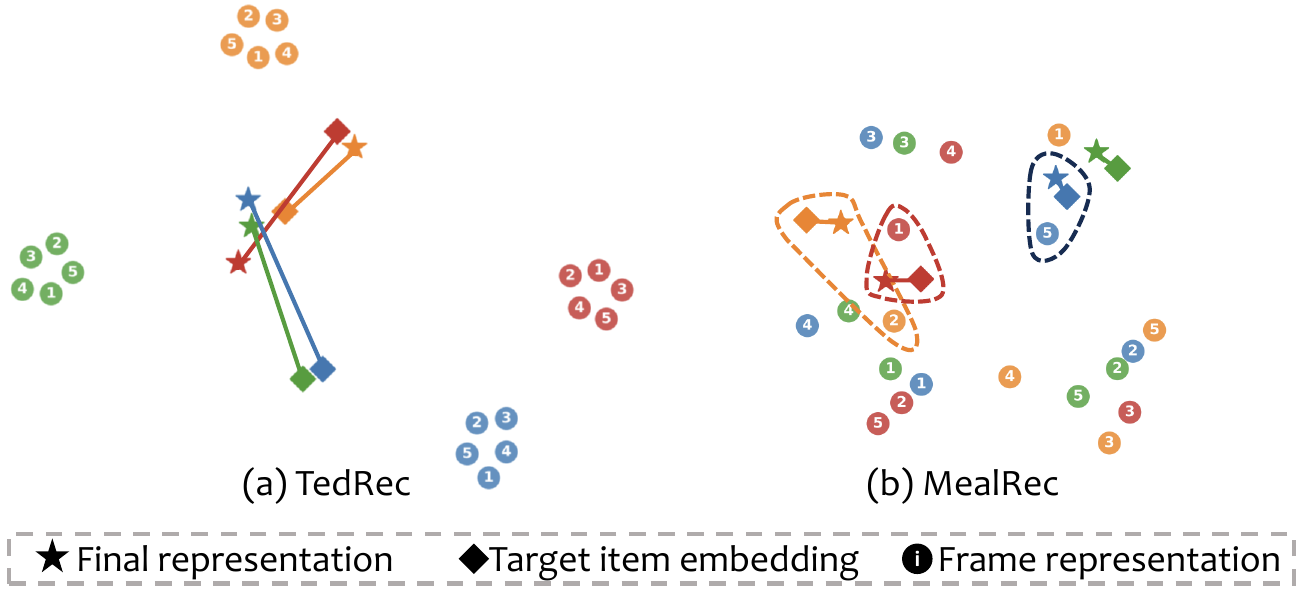}
    \vspace{-0.3cm}
    \caption{Visualization of the distribution on (a) TedRec and (b) \modelwospace. We utilize colors to differentiate users while using shapes and numbers to distinguish preference representation, target video embedding and frame representations within the same video, respectively.}
    \vspace{-0.5cm}
    \label{fig:Visualization}
\end{figure}

\subsubsection{Complexity Comparison}
As illustrated in Figure~\ref{tab:complexity_analysis}, the complexity comparison reveals a clear trade-off between recommendation performance and computational cost. Although IISAN-Versa delivers notable improvements over TedRec, it incurs substantial computational overhead, increasing both runtime and memory consumption and thus limiting its practicality for large-scale deployment. This issue becomes even more pronounced on the larger MicroLens-big dataset. In contrast, \model achieves the best performance across datasets while maintaining efficiency comparable to the lightweight backbone. Compared with TedRec, MealRec increases training time and memory footprint only moderately, and its overhead remains far below that of IISAN-Versa. Overall, these results demonstrate that \model improves recommendation accuracy without compromising computational efficiency, striking a favorable balance between performance and practical deployability.

\subsection{Universality Analysis (RQ4)} 

\label{sec:UniversalityAnalysis}
Beyond TedRec, we further instantiate \model on two additional backbones (\ie SASRec and CL4SRec) to assess its universality.

As illustrated in Figure~\ref{fig:Universityr_Analysis}, \model consistently outperforms all backbones and ablation variants across all datasets, indicating strong universality under different architectures and data characteristics. We further observe that the contributions of TCD and NPD vary by dataset. Specifically, on the MicroLens datasets, TCD brings larger gains than NPD, suggesting that modeling fine-grained temporal dynamics provides more informative preference signals than denoising alone. Combining TCD and NPD yields additional improvements over either component individually, implying that the two modules offer complementary benefits. Finally, the improvements are more pronounced on larger datasets, indicating that the proposed distillation mechanisms benefit from richer supervision and are well suited to large-scale recommendation settings.

\subsection{Visualization Analysis (RQ3)}
\label{sec:Visualization}  

To evaluate the effectiveness of video representations in guiding preference modeling, we visualize the representations of TedRec and \model via t-SNE in Figure~\ref{fig:Visualization}. 

Compared with TedRec, \model yields a smaller distance between the final user representation and the target item embedding, suggesting improved preference–target alignment. Moreover, \model induces a preference-aware structure among frame representations by distinguishing user preferences across different frame features. Specifically, interest-relevant frames lie closer to the user representation, indicating a strong correlation between temporal frame modeling and user preferences and supporting the effectiveness of our method.

\section{Conclusion}
This work investigates micro-video recommendation from a multi-granularity sequential modeling perspective, where effective prediction requires jointly capturing intra-video temporal dynamics and inter-video preference evolution. We proposed \modelwospace, a hierarchical diffusion framework that addresses two key challenges in prior endeavors: preference-irrelevant video representation extraction and inherent modality conflicts. Specifically, Temporal-guided Content Diffusion (TCD) refined intra-video representations by incorporating frame-level temporal guidance and personalized collaborative signals, while Noise-unconditional Preference Denoising (NPD) performs information fusion via blind denoising to directly recover informative user preferences from corrupted states without timestep-conditioned $\epsilon$-prediction. Extensive experiments on four datasets verify its effectiveness, universality, and robustness. Future work includes advancing \model toward an end-to-end optimization with the visual encoder, as well as incorporating complementary user feedback and enriched semantic cues to strengthen preference elicitation.

\balance

\bibliographystyle{unsrt}
\bibliography{software}

@inproceedings{wei2019mmgcn,
  title={MMGCN: Multi-modal graph convolution network for personalized recommendation of micro-video},
  author={Wei, Yinwei and Wang, Xiang and Nie, Liqiang and He, Xiangnan and Hong, Richang and Chua, Tat-Seng},
  booktitle={Proceedings of the 27th ACM international conference on multimedia},
  pages={1437--1445},
  year={2019}
}

@inproceedings{pan2023understanding,
  title={Understanding and modeling passive-negative feedback for short-video sequential recommendation},
  author={Pan, Yunzhu and Gao, Chen and Chang, Jianxin and Niu, Yanan and Song, Yang and Gai, Kun and Jin, Depeng and Li, Yong},
  booktitle={Proceedings of the 17th ACM conference on recommender systems},
  pages={540--550},
  year={2023}
}

@inproceedings{he2025short,
  title={Short Video Segment-level User Dynamic Interests Modeling in Personalized Recommendation},
  author={He, Zhiyu and Ling, Zhixin and Li, Jiayu and Guo, Zhiqiang and Ma, Weizhi and Luo, Xinchen and Zhang, Min and Zhou, Guorui},
  booktitle={Proceedings of the 48th International ACM SIGIR Conference on Research and Development in Information Retrieval},
  pages={1880--1890},
  year={2025}
}

@inproceedings{chen2024multi,
  title={A Multi-modal Modeling Framework for Cold-start Short-video Recommendation},
  author={Chen, Gaode and Sun, Ruina and Jiang, Yuezihan and Cao, Jiangxia and Zhang, Qi and Lin, Jingjian and Li, Han and Gai, Kun and Zhang, Xinghua},
  booktitle={Proceedings of the 18th ACM Conference on Recommender Systems},
  pages={391--400},
  year={2024}
}

@inproceedings{ye2025harnessing,
  title={Harnessing multimodal large language models for multimodal sequential recommendation},
  author={Ye, Yuyang and Zheng, Zhi and Shen, Yishan and Wang, Tianshu and Zhang, Hengruo and Zhu, Peijun and Yu, Runlong and Zhang, Kai and Xiong, Hui},
  booktitle={Proceedings of the AAAI Conference on Artificial Intelligence},
  volume={39},
  number={12},
  pages={13069--13077},
  year={2025}
}

@inproceedings{shang2023learning,
  title={Learning fine-grained user interests for micro-video recommendation},
  author={Shang, Yu and Gao, Chen and Chen, Jiansheng and Jin, Depeng and Wang, Meng and Li, Yong},
  booktitle={Proceedings of the 46th international ACM SIGIR conference on research and development in information retrieval},
  pages={433--442},
  year={2023}
}

@inproceedings{rendle2010factorizing,
  title={Factorizing personalized markov chains for next-basket recommendation},
  author={Rendle, Steffen and Freudenthaler, Christoph and Schmidt-Thieme, Lars},
  booktitle={Proceedings of the 19th international conference on World wide web},
  pages={811--820},
  year={2010}
}

@article{hidasi2015session,
  title={Session-based recommendations with recurrent neural networks},
  author={Hidasi, Bal{\'a}zs and Karatzoglou, Alexandros and Baltrunas, Linas and Tikk, Domonkos},
  journal={arXiv preprint arXiv:1511.06939},
  year={2015}
}

@inproceedings{kang2018self,
  title={Self-attentive sequential recommendation},
  author={Kang, Wang-Cheng and McAuley, Julian},
  booktitle={2018 IEEE international conference on data mining (ICDM)},
  pages={197--206},
  year={2018},
  organization={IEEE}
}

@inproceedings{sun2019bert4rec,
  title={BERT4Rec: Sequential recommendation with bidirectional encoder representations from transformer},
  author={Sun, Fei and Liu, Jun and Wu, Jian and Pei, Changhua and Lin, Xiao and Ou, Wenwu and Jiang, Peng},
  booktitle={Proceedings of the 28th ACM international conference on information and knowledge management},
  pages={1441--1450},
  year={2019}
}

@inproceedings{li2020time,
  title={Time interval aware self-attention for sequential recommendation},
  author={Li, Jiacheng and Wang, Yujie and McAuley, Julian},
  booktitle={Proceedings of the 13th international conference on web search and data mining},
  pages={322--330},
  year={2020}
}

@inproceedings{wu2020sse,
  title={SSE-PT: Sequential recommendation via personalized transformer},
  author={Wu, Liwei and Li, Shuqing and Hsieh, Cho-Jui and Sharpnack, James},
  booktitle={Proceedings of the 14th ACM conference on recommender systems},
  pages={328--337},
  year={2020}
}

@inproceedings{hu2023adaptive,
  title={Adaptive multi-modalities fusion in sequential recommendation systems},
  author={Hu, Hengchang and Guo, Wei and Liu, Yong and Kan, Min-Yen},
  booktitle={Proceedings of the 32nd ACM International Conference on Information and Knowledge Management},
  pages={843--853},
  year={2023}
}

@article{zhang2024multimodal,
  title={Multimodal pre-training for sequential recommendation via contrastive learning},
  author={Zhang, Lingzi and Zhou, Xin and Zeng, Zhiwei and Shen, Zhiqi},
  journal={ACM Transactions on Recommender Systems},
  volume={3},
  number={1},
  pages={1--23},
  year={2024},
  publisher={ACM New York, NY}
}

@inproceedings{wang2023diffusion,
  title={Diffusion recommender model},
  author={Wang, Wenjie and Xu, Yiyan and Feng, Fuli and Lin, Xinyu and He, Xiangnan and Chua, Tat-Seng},
  booktitle={Proceedings of the 46th international ACM SIGIR conference on research and development in information retrieval},
  pages={832--841},
  year={2023}
}

@inproceedings{hou2024collaborative,
  title={Collaborative filtering based on diffusion models: Unveiling the potential of high-order connectivity},
  author={Hou, Yu and Park, Jin-Duk and Shin, Won-Yong},
  booktitle={Proceedings of the 47th International ACM SIGIR Conference on Research and Development in Information Retrieval},
  pages={1360--1369},
  year={2024}
}

@inproceedings{zhu2024graph,
  title={Graph signal diffusion model for collaborative filtering},
  author={Zhu, Yunqin and Wang, Chao and Zhang, Qi and Xiong, Hui},
  booktitle={Proceedings of the 47th International ACM SIGIR Conference on Research and Development in Information Retrieval},
  pages={1380--1390},
  year={2024}
}

@article{li2023diffurec,
  title={Diffurec: A diffusion model for sequential recommendation},
  author={Li, Zihao and Sun, Aixin and Li, Chenliang},
  journal={ACM Transactions on Information Systems},
  volume={42},
  number={3},
  pages={1--28},
  year={2023},
  publisher={ACM New York, NY}
}

@inproceedings{liu2023diffusion,
  title={Diffusion augmentation for sequential recommendation},
  author={Liu, Qidong and Yan, Fan and Zhao, Xiangyu and Du, Zhaocheng and Guo, Huifeng and Tang, Ruiming and Tian, Feng},
  booktitle={Proceedings of the 32nd ACM International conference on information and knowledge management},
  pages={1576--1586},
  year={2023}
}

@inproceedings{ma2024plug,
  title={Plug-in diffusion model for sequential recommendation},
  author={Ma, Haokai and Xie, Ruobing and Meng, Lei and Chen, Xin and Zhang, Xu and Lin, Leyu and Kang, Zhanhui},
  booktitle={Proceedings of the AAAI conference on artificial intelligence},
  volume={38},
  number={8},
  pages={8886--8894},
  year={2024}
}

@inproceedings{ni2025content,
  title={A content-driven micro-video recommendation dataset at scale},
  author={Ni, Yongxin and Cheng, Yu and Liu, Xiangyan and Fu, Junchen and Li, Youhua and He, Xiangnan and Zhang, Yongfeng and Yuan, Fajie},
  booktitle={Proceedings of the 34th ACM International Conference on Information and Knowledge Management},
  pages={6486--6491},
  year={2025}
}

@inproceedings{xu2025mutual,
  title={Mutual Information-aware Knowledge Distillation for Short Video Recommendation},
  author={Xu, Han and Pan, Taoxing and Liu, Zhiqiang and Xu, Xiaoxiao},
  booktitle={Proceedings of the 31st ACM SIGKDD Conference on Knowledge Discovery and Data Mining V. 1},
  pages={2725--2734},
  year={2025}
}

@inproceedings{zhao2025multi,
  title={Multi-Granularity Distribution Modeling for Video Watch Time Prediction via Exponential-Gaussian Mixture Network},
  author={Zhao, Xu and Ma, Ruibo and Chen, Jiaqi and Zhao, Weiqi and Yang, Ping and Hu, Yao},
  booktitle={Proceedings of the Nineteenth ACM Conference on Recommender Systems},
  pages={309--318},
  year={2025}
}

@inproceedings{cui2025multi,
  title={Multi-modal multi-behavior sequential recommendation with conditional diffusion-based feature denoising},
  author={Cui, Xiaoxi and Lu, Weihai and Tong, Yu and Li, Yiheng and Zhao, Zhejun},
  booktitle={Proceedings of the 48th International ACM SIGIR Conference on Research and Development in Information Retrieval},
  pages={1593--1602},
  year={2025}
}

@article{tong2022videomae,
  title={Videomae: Masked autoencoders are data-efficient learners for self-supervised video pre-training},
  author={Tong, Zhan and Song, Yibing and Wang, Jue and Wang, Limin},
  journal={Advances in neural information processing systems},
  volume={35},
  pages={10078--10093},
  year={2022}
}

@article{fu2025efficient,
  title={Efficient and effective adaptation of multimodal foundation models in sequential recommendation},
  author={Fu, Junchen and Ge, Xuri and Xin, Xin and Karatzoglou, Alexandros and Arapakis, Ioannis and Zheng, Kaiwen and Ni, Yongxin and Joemon, Joemon M Jose},
  journal={IEEE Transactions on Knowledge and Data Engineering},
  year={2025},
  publisher={IEEE}
}

@article{sun2025noise,
  title={Is Noise Conditioning Necessary for Denoising Generative Models?},
  author={Sun, Qiao and Jiang, Zhicheng and Zhao, Hanhong and He, Kaiming},
  journal={arXiv preprint arXiv:2502.13129},
  year={2025}
}

@article{ni2023content,
  title={A Content-Driven Micro-Video Recommendation Dataset at Scale},
  author={Ni, Yongxin and Cheng, Yu and Liu, Xiangyan and Fu, Junchen and Li, Youhua and He, Xiangnan and Zhang, Yongfeng and Yuan, Fajie},
  journal={arXiv preprint arXiv:2309.15379},
  year={2023}
}

@inproceedings{shang2025large,
  title={A Large-scale Dataset with Behavior, Attributes, and Content of Mobile Short-video Platform},
  author={Shang, Yu and Gao, Chen and Li, Nian and Li, Yong},
  booktitle={Companion Proceedings of the ACM on Web Conference 2025},
  pages={793--796},
  year={2025}
}

@inproceedings{xie2022contrastive,
  title={Contrastive learning for sequential recommendation},
  author={Xie, Xu and Sun, Fei and Liu, Zhaoyang and Wu, Shiwen and Gao, Jinyang and Zhang, Jiandong and Ding, Bolin and Cui, Bin},
  booktitle={2022 IEEE 38th international conference on data engineering (ICDE)},
  pages={1259--1273},
  year={2022},
  organization={IEEE}
}

@inproceedings{zhang2024ssdrec,
  title={Ssdrec: Self-augmented sequence denoising for sequential recommendation},
  author={Zhang, Chi and Han, Qilong and Chen, Rui and Zhao, Xiangyu and Tang, Peng and Song, Hongtao},
  booktitle={2024 IEEE 40th International Conference on Data Engineering (ICDE)},
  pages={803--815},
  year={2024},
  organization={IEEE}
}

@inproceedings{mao2025distinguished,
  title={Distinguished quantized guidance for diffusion-based sequence recommendation},
  author={Mao, Wenyu and Liu, Shuchang and Liu, Haoyang and Liu, Haozhe and Li, Xiang and Hu, Lantao},
  booktitle={Proceedings of the ACM on Web Conference 2025},
  pages={425--435},
  year={2025}
}

@inproceedings{yuan2023go,
  title={Where to go next for recommender systems? id-vs. modality-based recommender models revisited},
  author={Yuan, Zheng and Yuan, Fajie and Song, Yu and Li, Youhua and Fu, Junchen and Yang, Fei and Pan, Yunzhu and Ni, Yongxin},
  booktitle={Proceedings of the 46th International ACM SIGIR Conference on Research and Development in Information Retrieval},
  pages={2639--2649},
  year={2023}
}

@inproceedings{fu2024iisan,
  title={IISAN: Efficiently Adapting Multimodal Representation for Sequential Recommendation with Decoupled PEFT},
  author={Fu, Junchen and Ge, Xuri and Xin, Xin and Karatzoglou, Alexandros and Arapakis, Ioannis and Wang, Jie and Jose, Joemon M},
  booktitle={Proceedings of the 47th International ACM SIGIR Conference on Research and Development in Information Retrieval},
  pages={687--697},
  year={2024}
}

@inproceedings{xu2024tedrec,
  author    = {Lanling Xu and Zhen Tian and Bingqian Li and Junjie Zhang and Daoyuan Wang and Hongyu Wang and Jinpeng Wang and Sheng Chen and Wayne Xin Zhao},
  title     = {Sequence-level Semantic Representation Fusion for Recommender Systems},
  booktitle = {{CIKM}},
  publisher = {{ACM}},
  year      = {2024}
}

@inproceedings{lu2025dmmd4sr,
  title={Dmmd4sr: Diffusion model-based multi-level multimodal denoising for sequential recommendation},
  author={Lu, Weihai and Yin, Li},
  booktitle={Proceedings of the 33rd ACM International Conference on Multimedia},
  pages={6363--6372},
  year={2025}
}

@inproceedings{devlin2019bert,
  title={Bert: Pre-training of deep bidirectional transformers for language understanding},
  author={Devlin, Jacob and Chang, Ming-Wei and Lee, Kenton and Toutanova, Kristina},
  booktitle={Proceedings of the 2019 conference of the North American chapter of the association for computational linguistics: human language technologies, volume 1 (long and short papers)},
  pages={4171--4186},
  year={2019}
}

@inproceedings{pennington2014glove,
  title={Glove: Global vectors for word representation},
  author={Pennington, Jeffrey and Socher, Richard and Manning, Christopher D},
  booktitle={Proceedings of the 2014 conference on empirical methods in natural language processing (EMNLP)},
  pages={1532--1543},
  year={2014}
}

@article{ho2022video,
  title={Video diffusion models},
  author={Ho, Jonathan and Salimans, Tim and Gritsenko, Alexey and Chan, William and Norouzi, Mohammad and Fleet, David J},
  journal={Advances in neural information processing systems},
  volume={35},
  pages={8633--8646},
  year={2022}
}

@article{zhao2025nextquill,
  title={NextQuill: Causal Preference Modeling for Enhancing LLM Personalization},
  author={Zhao, Xiaoyan and You, Juntao and Zhang, Yang and Wang, Wenjie and Cheng, Hong and Feng, Fuli and Ng, See-Kiong and Chua, Tat-Seng},
  journal={ICLR},
  year={2026}
}

@article{zhang2025reinforced,
  title={Reinforced Latent Reasoning for LLM-based Recommendation},
  author={Zhang, Yang and Xu, Wenxin and Zhao, Xiaoyan and Wang, Wenjie and Feng, Fuli and He, Xiangnan and Chua, Tat-Seng},
  journal={ICLR},
  year={2026}
}

@article{zhao2025reinforced,
  title={Reinforced Strategy Optimization for Conversational Recommender Systems via Network-of-Experts},
  author={Zhao, Xiaoyan and Yan, Ming and Zhang, Yang and Deng, Yang and Wang, Jian and Zhu, Fengbin and Qiu, Yilun and Cheng, Hong and Chua, Tat-Seng},
  journal={arXiv preprint arXiv:2509.26093},
  year={2025}
}

@article{TriCDR,
author = {Ma, Haokai and Xie, Ruobing and Meng, Lei and Chen, Xin and Zhang, Xu and Lin, Leyu and Zhou, Jie},
title = {Triple Sequence Learning for Cross-domain Recommendation},
year = {2024},
publisher = {Association for Computing Machinery},
address = {New York, NY, USA},
volume = {42},
number = {4},
issn = {1046-8188},
url = {https://doi.org/10.1145/3638351},
doi = {10.1145/3638351},
journal = {ACM Trans. Inf. Syst.},
}

@article{NS4RS,
		author = {Ma, Haokai and Xie, Ruobing and Meng, Lei and Feng, Fuli and Du, Xiaoyu and Sun, Xingwu and Kang, Zhanhui and Meng, Xiangxu},
		title = {Negative Sampling in Recommendation: A Survey and Future Directions},
		year = {2026},
		publisher = {Association for Computing Machinery},
		address = {New York, NY, USA},
		url = {https://doi.org/10.1145/3793855},
		doi = {10.1145/3793855},
		journal = {ACM Trans. Inf. Syst.},
}

@inproceedings{SeeDRec,
  author       = {Haokai Ma and
                  Ruobing Xie and
                  Lei Meng and
                  Yimeng Yang and
                  Xingwu Sun and
                  Zhanhui Kang},
  title        = {SeeDRec: Sememe-based Diffusion for Sequential Recommendation},
  booktitle    = {Proceedings of the Thirty-Third International Joint Conference on Artificial Intelligence, {IJCAI} 2024, Jeju, South Korea, August 3-9, 2024},
  pages        = {2270--2278},
  year         = {2024}
}

@inproceedings{RealHNS,
  author       = {Haokai Ma and
                  Ruobing Xie and
                  Lei Meng and
                  Xin Chen and
                  Xu Zhang and
                  Leyu Lin and
                  Jie Zhou},
  title        = {Exploring False Hard Negative Sample in Cross-Domain Recommendation},
  booktitle    = {Proceedings of the 17th {ACM} Conference on Recommender Systems, RecSys 2023, Singapore, Singapore, September 18-22, 2023},
  pages        = {502--514},
  year         = {2023},
  doi          = {10.1145/3604915.3608791}
}

@inproceedings{MCDRec,
author = {Ma, Haokai and Yang, Yimeng and Meng, Lei and Xie, Ruobing and Meng, Xiangxu},
title = {Multimodal Conditioned Diffusion Model for Recommendation},
year = {2024},
publisher = {Association for Computing Machinery},
address = {New York, NY, USA},
url = {https://doi.org/10.1145/3589335.3651956},
doi = {10.1145/3589335.3651956},
booktitle = {Companion Proceedings of the ACM Web Conference 2024},
pages = {1733–1740},
numpages = {8},
location = {Singapore, Singapore},
series = {WWW '24}
}

@misc{HorizonRec,
      title={Align-for-Fusion: Harmonizing Triple Preferences via Dual-oriented Diffusion for Cross-domain Sequential Recommendation}, 
      author={Yongfu Zha and Xinxin Dong and Haokai Ma and Yonghui Yang and Xiaodong Wang},
      year={2025},
      eprint={2508.05074},
      archivePrefix={arXiv},
      primaryClass={cs.IR},
      url={https://arxiv.org/abs/2508.05074}, 
}

@inproceedings{CIERec,
  author       = {Haokai Ma and
                  Zhuang Qi and
                  Xinxin Dong and
                  Xiangxian Li and
                  Yuze Zheng and
                  Xiangxu Meng and
                  Lei Meng},
  title        = {Cross-Modal Content Inference and Feature Enrichment for Cold-Start Recommendation},
  booktitle    = {International Joint Conference on Neural Networks, {IJCNN} 2023, Gold Coast, Australia, June 18-23, 2023},
  pages        = {1--8},
  publisher    = {{IEEE}},
  year         = {2023},
  doi          = {10.1109/IJCNN54540.2023.10191979}
}

\end{sloppypar}
\end{document}